# Spulen und Transformatoren – häufig benutzt und selten richtig erklärt

## Michael Lenz*, Torsten Schmidt+

*Technische Universität Dresden, Fak. Elektrotechnik und Informationstechnik, Institut für Festkörper-
elektronik, D-01062 Dresden, michael.lenz@tu-dresden.de
+Technische Universität Dresden, Fak. Elektrotechnik und Informationstechnik, Institut für Grundlagen der
Elektrotechnik, D-01062 Dresden, torsten.schmidt1@tu-dresden.de

**Kurzfassung**

Die elektrotechnischen Bauelemente „Spule" und „Transformator" werden in vielen einführenden Lehrbüchern der Physik, der Elektrotechnik, in Schulbüchern und in Anfängerpraktika an Hochschulen behandelt. Viele Darstellungen zeichnen jedoch ein unverständliches oder teilweise sogar falsches Bild der zugrundeliegenden Physik und vermitteln so zahlreiche Fehlvorstellungen. Dazu gehören:

- Primär- und Sekundärspannung an einem Transformator haben unterschiedliche Vorzeichen.
- Die Funktionsweise des Transformators basiert auf elektromagnetischer Induktion allein.
- Bei Spule und Transformator kompensieren sich die angelegte Klemmenspannung und die induzierte Spannung.
- Transformatorkerne dienen als magnetischer Zwischenspeicher für die im Transformator übertragene Energie. Die Energie fließt durch den Transformatorkern.
- Streufelder beschreiben Nebensächlichkeiten und tragen beim Transformator nicht zur Leistungsübertragung bei.
- Je größer der Laststrom, umso eher sättigt der Transformatorkern.
- Eine hohe Windungszahl an der Primärseite eines Transformators führt zu einem großen magnetischen Fluss im Kern
- Bei Transformatoren mit Luftspalt verringert sich der Kopplungsfaktor, weil der Luftspalt die Streuinduktivität vergrößert

Im Artikel werden zunächst die grundlegenden Eigenschaften von Spule und Transformator unter Annahme idealtypischer Verhältnisse aus den Maxwellgleichungen abgeleitet und anschließend die genannten fehlerhaften Darstellungen diskutiert.

## 1. Einleitung

Ist es erforderlich, knapp 150 Jahre nach Veröffentlichung von J. Clerk Maxwells Feldtheorie [1] und mehr als 100 Jahre nach der Veröffentlichung von Gisbert Kapps Monographie über Transformatoren [2] überhaupt noch eine Einführung in die Transformatorphysik zu schreiben?

Aus Sicht der Autoren trifft dies zu, denn die einführende technisch-physikalische Literatur befindet sich teilweise in einem bedauerlich schlechten Zustand. So weisen viele Darstellungen Inkonsistenzen und logische Widersprüche auf, während kaum ein Werk die typischen Grundfragen jedes neugierigen Schülers oder Studenten schlüssig beantworten kann. Somit sind diese Bauelemente zwar in der technischen Literatur und Anwendung allgegenwärtig, bleiben für viele Schüler und Studenten jedoch grundlegend unverständlich.

Mit der vorliegenden Veröffentlichung soll auf die Notwendigkeit einer Überarbeitung hingewiesen werden. Zudem werden gleichfalls konkrete Hinweise zu einer verständlichen Darstellung gegeben. In diesem Artikel geht es im Wesentlich um die klare Ableitung der Transformatorgleichungen für den idealen Transformator aus den Maxwell-Gleichungen. Gleichermaßen lassen sich die Inhalte in etwas komplexeren Umfang auf reale Transformatoren mit Eisenverlusten, Streuinduktivitäten und ohmschen Leiterwiderständen übertragen, was allerdings nicht Inhalt dieses Aufsatzes sein soll. Neben der Anwendung der Maxwell-Gleichungen auf den idealen Transformator diskutieren wir ebenfalls die auffälligsten Mängel der geläufigen Darstellungen.

## 2. Vorzeichenkonventionen

Da ein wichtiger Teil der Erläuterungen die Behandlung der Vorzeichen betrifft, soll einleitend das Konzept der sogenannten Zählpfeile beschrieben werden. Hierbei handelt es sich um Pfeile, die die Bezugsrichtung für Netzwerkgrößen wie elektrische Ströme und Spannungen definieren. In anderen Zusammenhängen eignen sich Zählpfeile aber auch zur Darstellung nichtelektrischer Größen wie mechanischer Kräfte oder Auslenkungen.

Praktisch läuft die Definition der Zählpfeile auf eine Vereinbarung darüber hinaus, in welcher Richtung das jeweilige zur Messung der physikalischen Größe verwendete Messgerät in die Anordnung einzubauen ist. Die Zählpfeile ermöglichen es auf diese Weise, physikalische Zusammenhänge vorzeichenrichtig zu notieren, ohne die tatsächlichen Richtungen der Feldgrößen im Vorhinein zu kennen. Dies ist insbesondere in umfangreichen Zusammenschaltungen verschiedener Bauteile notwendig, bei denen sich die Vorzeichen sonst nur sehr mühsam erfassen ließen.

### 2.1. Spannungszählpfeile

Die elektrische Spannung zwischen zwei Punkten P und Q entlang einer vorgegebenen Kurve $s$ ist allgemein durch das Linienintegral

$$u(t) = \int_P^Q \vec{E} \cdot d\vec{s} \qquad \{1\}$$

über die elektrische Feldstärke $E$ definiert. Um die Integrationsrichtung (und damit verbunden das Vorzeichen der Spannung) zu veranschaulichen, werden elektrische Schaltbilder typischerweise mit einem Spannungszählpfeil versehen, der vom Anfangs- zum Endpunkt der Kurve zeigt (**Abb. 1**).

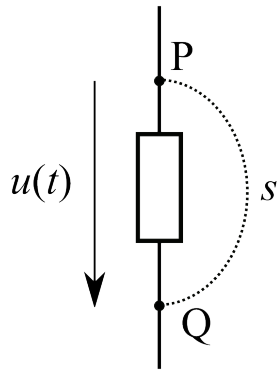

**Abb. 1:** Der eingezeichnete Spannungszählpfeil legt fest, dass beim Bilden des Linienintegrales über die elektrische Feldstärke $E$ die Kurve $s$ im Punkt P beginnt und im Punkt Q endet.

Eine besondere Bedeutung kommt dem Integralwert im Falle von Potentialfeldern zu. Für solche ist der Integralwert nicht vom durchlaufenen Weg, sondern ausschließlich vom Anfangs- und Endpunkten abhängig. Beim Vorliegen von Induktion ist das Integral hingegen im Allgemeinen wegabhängig und die Definition der Spannung zwischen den beiden Punkten P und Q somit mehrdeutig.

Wenn im Laufe des vorliegenden Artikels von „Klemmenspannungen" beispielsweise einer Spule gesprochen wird, so sind solche Verbindungen zwischen den Klemmen der Spule gemeint, die weit entfernt vom magnetischen Kern verlaufen und keine zusätzlichen Wirbel des E-Feldes einschließen.

### 2.2. Stromzählpfeile

Die verallgemeinerte elektrische Stromstärke lässt sich allgemein über die Summe aus der Leiterstromdichte $\vec{j}(t, \vec{x})$ und der zeitlichen Ableitung der elektrischen Verschiebungsdichte $\vec{D}(t, \vec{x})$, jeweils integriert über die Fläche $A$, in der Form

$$i(t) = \int_A \left[ \vec{j}(t, \vec{x}) + \frac{\partial \vec{D}(t, \vec{x})}{\partial t} \right] \cdot d\vec{A} \qquad \{2\}$$

darstellen [12].

Der in **Abb. 2**, links, dargestellte *Stromzählpfeil* definiert dabei die Orientierung der Fläche $A$. Für ebene Flächen gilt, dass der Stromzählpfeil die Richtung des zur Querschnittsfläche zugehörigen Normalenvektors angibt. Im vorliegenden Beispiel zeigt der Stromzählpfeil nach unten. Das bedeutet, dass der Stromzählpfeil mit einer positiven Zahl versehen werden muss, wenn ein über positive Ladungsträger definierter technischer Strom „von oben nach unten" fließt.

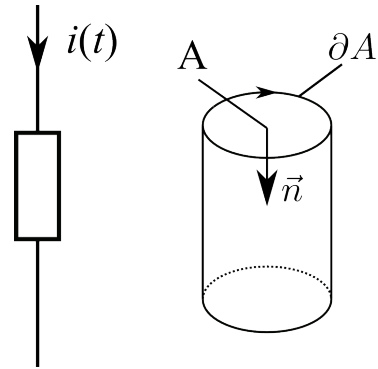

**Abb. 2:** Das rechte Teilbild zeigt eine Vergrößerung des Leiters. Der zum Strom $i(t)$ eingezeichnete Stromzählpfeil kennzeichnet die Orientierung der Fläche $A$, über die die Stromdichte $j$ integriert wird. Die Orientierung wird durch den zugehörigen Normalenvektor $\vec{n}$ gekennzeichnet.

Stellt man die rechte Seite von Gleichung {2} mithilfe des Durchflutungssatzes durch die magnetische Feldstärke $\vec{H}$ in der Form

$$\oint_{\partial A} \vec{H}(t, \vec{x}) \cdot d\vec{s} = \int_A \left[ \vec{j}(t, \vec{x}) + \frac{\partial \vec{D}(t, \vec{x})}{\partial t} \right] \cdot d\vec{A} \; \{3\}$$

dar, so zeigt sich, dass alle Flächen mit gleicher Konturlinie den gleichen Strom führen. Die Bezugsrichtung des Stromes lässt sich dementsprechend auch über die Umlaufrichtung des Ringintegrals um den Flächenrand definieren. Allgemein gilt dabei die Konvention, dass der Bezugszählpfeil des Stromes (d. h. die Orientierung der Fläche) und die Randlinie der Fläche *rechtshändig* zueinander zugeordnet sind. Zeigt der Daumen der rechten Hand in Richtung des Stromzählpfeiles, so ist das Linienintegral über das $H$-Feld in Richtung der übrigen Finger der rechten Hand zu bestimmen.

## 3. Idealisierte Transformatorgleichungen

### 3.1. Transformatormodell

Zur Darstellung der grundlegenden Transformationseigenschaften soll von einem einphasigen Transformator entsprechend **Abb. 3** ausgegangen werden, wobei die folgenden Bedingungen gelten sollen:

- Transformatorkern mit hoher Permeabilität $\mu_r \to \infty$ und linearem Materialverhalten,

- ideale Leitfähigkeit $\kappa \to \infty$ der elektrischen Leitungen und

- Vernachlässigung der Streufelder bei der Berechnung des magnetischen Flusses.

Das Ziel besteht darin, die für das grundlegende Verständnis wesentlichen Zusammenhänge abzubilden. Aus diesem Grund wird eine Darstellung gewählt, bei der die Materialparameter erst möglichst spät verwendet werden. Dies erleichtert insbesondere das Verständnis des Transformatorverhaltens bei Sättigung.

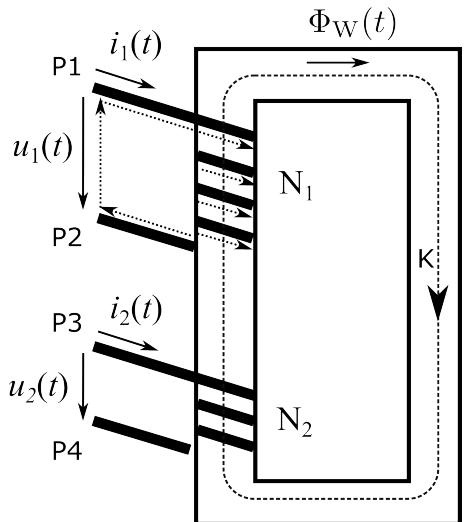

**Abb. 3:** Verwendetes Modell eines einphasigen Transformators mit Kern.

### 3.2. Spannungstransformation

#### 3.2.1. Induktionsgesetz

Zur Herleitung der Spannungstransformation wird das Induktionsgesetz verwendet. Dieses lautet in differentieller Form:

$$\nabla \times \vec{E} = -\frac{\partial \vec{B}}{\partial t} \qquad \{4\}$$

Setzt man das differentielle Induktionsgesetz in den Satz von Stokes

$$\oint_{\partial A} \vec{E} \cdot \mathrm{d}\vec{s} = \int_{A} (\vec{\nabla} \times \vec{E}) \cdot \mathrm{d}\vec{A} \qquad \{5\}$$

ein, so ergibt sich das Induktionsgesetz in Integralform

$$\oint_{\partial A} \vec{E} \cdot \mathrm{d}\vec{s} = -\int_{A} \frac{\partial \vec{B}}{\partial t} \cdot \mathrm{d}\vec{A} \qquad \{6\}$$

wobei das Ringintegral über die elektrische Feldstärke im Rahmen dieses Aufsatzes als induzierte Spannung bezeichnet wird[1]:

$$u_{\mathrm{ind}} := \oint_{\partial A} \vec{E} \cdot \mathrm{d}\vec{s} \qquad \{7\}$$

Unter der Voraussetzung hinreichend glatter Vektorfelder und orientierbarer Flächen (die im Rahmen dieses Artikels grundsätzlich vorausgesetzt werden sollen), ist das Induktionsgesetz in der Form nach Gleichung {6} mathematisch äquivalent mit dem Induktionsgesetz in Differentialform und lässt sich entgegen vielen gegenteiligen Aussagen in der Literatur problemlos auch auf bewegte Objekte anwenden (vgl. [3], [4], [5]). Dabei ist jedoch streng zwischen der Bewegung der mathematischen Kurve $\partial A$ und der Bewegung der physikalischen Objekte (Leiterschleife, Ladungen) zu unterscheiden.

Um zu einer Darstellung des Induktionsgesetzes zu gelangen, die die Ableitung des magnetischen Flusses enthält, wird auf beiden Seiten von Gleichung {6} der Term

$$\oint_{\partial A} (\vec{u} \times \vec{B}) \cdot \mathrm{d}\vec{s} \qquad \{8\}$$

addiert. Hierbei bezeichnet $\vec{u}$ die lokale Geschwindigkeit der Konturlinie. Es ergibt sich:

$$\oint_{\partial A} (\vec{E} + \vec{u} \times \vec{B}) \cdot \mathrm{d}\vec{s} =$$
$$-\int_{A} \frac{\partial \vec{B}}{\partial t} \cdot \mathrm{d}\vec{A} + \oint_{\partial A} (\vec{u} \times \vec{B}) \cdot \mathrm{d}\vec{s} \qquad \{9\}$$

Setzt man die Maxwellgleichung $\nabla \cdot \vec{B} = 0$ (Nichtvorhandensein magnetischer Ladungen) in die mathematische Identität [6]

$$\frac{\mathrm{d}}{\mathrm{d}t} \int_{A} \vec{B} \cdot \mathrm{d}\vec{A} = \int_{A} (\nabla \cdot \vec{B}) \cdot \vec{u} \cdot \mathrm{d}\vec{A} -$$
$$\oint_{\partial A} (\vec{u} \times \vec{B}) \cdot \mathrm{d}\vec{s} + \int_{A} \frac{\partial \vec{B}}{\partial t} \cdot \mathrm{d}\vec{A} \qquad \{10\}$$

ein, so zeigt es sich, dass die rechte Seite von Gleichung {9} gerade die negative zeitliche Ableitung des magnetischen Flusses ist. Das Induktionsgesetz kann also in voller Allgemeingültigkeit auch durch

$$\oint_{\partial A} (\vec{E} + \vec{u} \times \vec{B}) \cdot \mathrm{d}\vec{s} = -\frac{\mathrm{d}}{\mathrm{d}t} \int_{A} \vec{B} \cdot \mathrm{d}\vec{A} \qquad \{11\}$$

beschrieben werden.

Unter der Voraussetzung, dass die Kurve $\partial A$ keine zeitlichen Änderungen aufweist ($u = 0$), vereinfacht sich das Induktionsgesetz zu

---

[1] Manche Autoren verstehen unter der induzierten Spannung abweichend von der hier verwendeten Definition die Klemmenspannung an einer Leiterschleife. Bei bewegten Flächenkonturen weichen beide Definitionen leider i. A. voneinander ab. Im Rahmen dieses Artikels besteht jedoch keine Verwechselungsgefahr, da ausschließlich zeitlich unveränderliche Flächenkonturen auftreten.

$$\oint_{\partial A} \vec{E} \cdot \mathrm{d}\vec{s} = -\frac{\mathrm{d}}{\mathrm{d}t} \int_A \vec{B} \cdot \mathrm{d}\vec{A} = -\frac{\mathrm{d}\Phi}{\mathrm{d}t}, \qquad \{12\}$$

wobei der magnetische Fluss $\Phi$ wie üblich als Flächenintegral über die magnetische Flussdichte in der Form

$$\Phi(t) := \int_A \vec{B} \cdot \mathrm{d}\vec{A} \qquad \{13\}$$

verwendet wird.

Da im Rahmen des vorliegenden Aufsatzes ausschließlich ruhende Flächenkonturen betrachtet werden, wird im Folgenden das Induktionsgesetz in der nicht-allgemeingültigen Form entsprechend Gleichung {12} verwendet.

### 3.2.2. Spulenfläche

Als Integrationsweg für die induzierte Spannung an der Primär- und Sekundärspule des Transformators soll der Umlaufweg durch den Wickeldraht betrachtet werden, der durch eine kurze „Luftstrecke" zwischen den Anschlussklemmen zu einem geschlossenen Umlaufweg vervollständigt wird. Dieser ist in **Abb. 3** durch die gestrichelten Pfeile neben dem Wickeldraht gekennzeichnet.

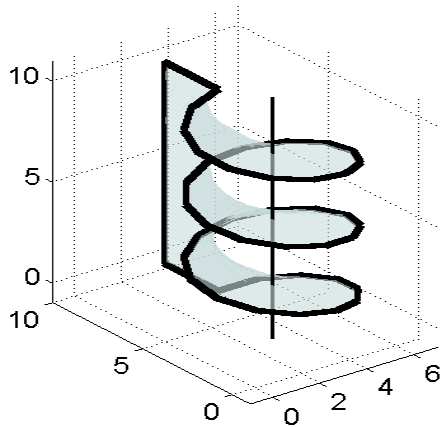

**Abb. 4:** Der Wickeldraht einer Spule kann zusammen mit der Verbindungslinie zwischen den Klemmen als Randkurve einer zusammenhängenden Fläche angesehen werden (vgl. auch [7]). Es ist zu erkennen, dass die Feldlinien die Fläche im Bereich des Magnetkernes mehrfach durchstoßen.

**Abb. 4** zeigt anhand eines Beispiels, wie man sich eine zu einer solchen Wicklung zugehörige Induktionsfläche anschaulich vorstellen kann.

Die Induktionsfläche lässt sich in einer schulischen oder universitären Veranstaltung mit einfachen Mitteln durch ein Seifenblasenexperiment veranschaulichen. Dazu lässt man die Lernenden einen Draht um einen zylindrischen Körper (z. B. einen Besenstiel) wickeln und beide Drahtenden anschließend miteinander verbinden und in Seifenwasser tunken. Nach dem Herausziehen der Spule aus der Flüssig-keit zeichnet die Seifenhaut eine zur Berechnung des Spulenflusses geeignete Fläche nach (Minimalfläche) und kann von allen Seiten betrachtet werden.

Zu der üblichen Formulierung des Induktionsgesetzes mit den Windungszahlen $N_1$ und $N_2$ gelangt man schließlich, indem man die von der Primär- und Sekundärwicklung umfassten Flüsse $\Phi_1$ und $\Phi_2$ im Idealfall als Vielfaches des Flusses durch eine einzelne Windung ausdrückt:

$$\Phi_1 = N_1 \cdot \Phi_{\mathrm{W}}$$
$$\Phi_2 = N_2 \cdot \Phi_{\mathrm{W}}. \qquad \{14\}$$

Es sei daran erinnert, dass es bei der Berechnung des magnetischen Flusses ausschließlich auf die Randlinie ankommt, nicht aber auf die genaue Ausformung der Fläche. Der Grund ist, dass die Differenz zwischen den magnetischen Flüssen durch zwei Flächen $A_1$ und $A_2$ mit gleicher Randkurve und Flächenorientierung sich jeweils auf ein Integral über das von beiden Flächen eingeschlossenen Volumen $V$ zurückführen lässt, das wegen des Nichtvorhandenseins magnetischer Ladungen ($\nabla \cdot \vec{B} = 0$) stets den Wert Null aufweist:

$$\oint_{A_1} \vec{B} \cdot \mathrm{d}\vec{A} - \oint_{A_2} \vec{B} \cdot \mathrm{d}\vec{A} = \oint_{\partial V} \vec{B} \cdot \mathrm{d}\vec{A} =$$
$$\underset{\text{(Satz von Gauß)}}{=} \int_V (\nabla \cdot \vec{B}) \cdot \mathrm{d}V = 0 \qquad \{15\}$$

### 3.2.3. Transformationsgleichung

Wendet man das Induktionsgesetz in der beschriebenen Weise (**Abb. 3**) auf die an der Primärseite des Transformators vorhandene Spule an, so ergibt sich

$$\oint \vec{E}\mathrm{d}\vec{s} = \overbrace{\int_{P1}^{P2} \vec{E}\mathrm{d}\vec{s}}^{\text{Draht}} + \overbrace{\int_{P2}^{P1} \vec{E}\mathrm{d}\vec{s}}^{\text{Luft}} =$$
$$= R_{\mathrm{D1}} \cdot i_1(t) - u_1(t) = \qquad \{16\}$$
$$= -\frac{\mathrm{d}\Phi_1}{\mathrm{d}t} = -N_1 \frac{\mathrm{d}\Phi_{\mathrm{W}}}{\mathrm{d}t},$$

wobei $R_{\mathrm{D1}}$ den ohmschen Widerstand des Drahtes auf der Primärseite bezeichnet. Kann der Drahtwiderstand vernachlässigt werden, so folgt mit $R_{\mathrm{D1}} \to 0$ die Näherung:

$$u_1(t) = N_1 \frac{\mathrm{d}\Phi_{\mathrm{W}}}{\mathrm{d}t} \qquad \{17\}$$

Analog gilt für die Sekundärseite:

$$u_2(t) = N_2 \cdot \frac{\mathrm{d}\Phi_{\mathrm{W}}}{\mathrm{d}t} \qquad \{18\}$$

Durch Einsetzen von Gleichung {17} in Gleichung {18} ergibt sich unmittelbar die Gleichung für die Spannungstransformation am idealen Transformator:

$$\frac{u_1(t)}{N_1} = \frac{\mathrm{d}\Phi_{\mathrm{W}}}{\mathrm{d}t} = \frac{u_2(t)}{N_2}. \qquad \{19\}$$

Bemerkenswert ist, dass zur Herleitung dieser Gleichung ausschließlich die elektrischen Spannungen, nicht aber die Ströme betrachtet werden müssen.

### 3.2.4. Anmerkungen

Es sei angemerkt, dass die Induktionsspannung aufgrund der (vorwiegend im magnetischen Kern) vorhandenen Flussdichteänderung kein Potentialfeld ist. Das bedeutet, dass das Integral über die elektrische Feldstärke für einen Verbindungsweg zwischen zwei Punkten vom gewählten Integrationsweg abhängt. Das ist im Beispiel leicht daran zu erkennen, dass das Integral über die elektrische Feldstärke zwischen den Punkten P1 und P2 bei einer Verbindung durch den Draht näherungsweise gleich Null ist, während es – über den Luftweg integriert – einen nicht zu vernachlässigenden Zahlenwert aufweist. Dementsprechend kann man streng genommen nicht von den Klemmenspannungen $u_1(t)$ bzw. $u_2(t)$ sprechen kann, ohne gleichzeitig die zugehörige Integrationskurve anzugeben.

Für Integrationswege zwischen den Transformatorklemmen, die weit genug vom magnetischen Kern entfernt verlaufen, sind die Flussdichteänderungen jedoch meist so gering, dass die Abhängigkeit vom durchlaufenen Weg vernachlässigt werden kann. Somit kann hier das allgemeine Verständnis zur Spannung als Potenzialdifferenz aufrecht gehalten werden. Im vorliegenden Aufsatz wird somit (wie auch in der einschlägigen Literatur üblich) weiterhin von *der* Klemmenspannung gesprochen und dabei implizit vorausgesetzt, dass das elektrische Feld über Wege „weit außerhalb des magnetischen Feldes" integriert wird.

### 3.3. Stromtransformation

### 3.3.1. Durchflutungsgesetz

Zur Herleitung der Gleichung der Stromtransformation wird das Durchflutungsgesetz verwendet. Dieses lautet in differentieller Form:

$$\vec{\nabla} \times \vec{H} = \vec{j} + \frac{\partial \vec{D}}{\partial t} \qquad \{20\}$$

Die differentielle Form kann analog wie beim Induktionsgesetz mittels des Stokes'schen Satzes in die Integralform überführt werden:

$$\oint_{\partial A} \vec{H}\,\mathrm{d}\vec{s} = \int_A \left( \vec{j} + \frac{\partial \vec{D}}{\partial t} \right) \cdot \mathrm{d}\vec{A} = i_\mathrm{L} + i_\mathrm{V} \quad \{21\}$$

Mit $i_\mathrm{L}$ und $i_\mathrm{V}$ werden hierbei der Leiterstrom bzw. der sogenannte Verschiebungsstrom verstanden. Simonyi [8] weist darauf hin, dass die Verschiebungsströme bei belasteten 50 Hz Transformatoren zur Berechnung der Stromtransformation im Allgemeinen vernachlässigt werden können, da der Leiterstrom typischerweise um mehrere Größenordnungen größer ist als der Verschiebungsstrom $i_\mathrm{L} \gg i_\mathrm{V}$.

Eine Vernachlässigung des Verschiebungsstromes führt implizit dazu, dass Wellenausbreitungseffekte zwischen den beiden Transformatorseiten vernachlässigt werden. Es muss also bei Anwendung der vereinfachten Gleichungen grundsätzlich gewährleistet sein, dass die Linearabmessungen des Transformators klein im Vergleich zu der Wellenlänge sind. Das ist für 50 Hz-Transformatoren mit einer Wellenlänge der elektromagnetischen Welle von $\lambda = c/f = 6000$ km keine wesentliche Einschränkung, muss aber bei Hochfrequenztransformatoren insbesondere für die meist unerwünschten Oberwellen (mit vielfach kleinerer Wellenläng für die signalführenden Welle) im Einzelfall geprüft werden.

Das Durchflutungsgesetz ergibt sich somit in der vereinfachten Form zu:

$$\oint \vec{H} \cdot \mathrm{d}\vec{s} = \sum_k I_k \qquad \{22\}$$

Der üblichen Vorzeichenkonvention entsprechend sind die Zählpfeile dabei der Integrationsrichtung für das Linienintegral rechtshändig zugeordnet (**Abb. 5**).

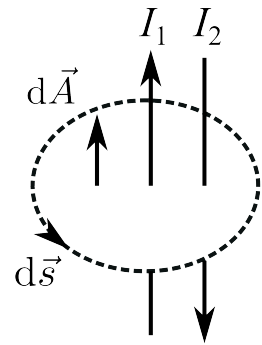

$$\oint \vec{H}\,\mathrm{d}\vec{s} = \int \vec{j}_\mathrm{L}\,\mathrm{d}\vec{A} = I_1 - I_2$$

**Abb. 5:** Vorzeichenkonvention bei der Anwendung des Durchflutungsgesetzes. $\vec{j}_\mathrm{L}$ ist die Leiterstromdichte.

Im Folgenden wird die magnetische Feldstärke $H$ im Magnetkern (**Abb. 3**) betrachtet: Wird das Transformator mit Wechselspannung betrieben, so gilt für den magnetischen Fluss im Kern[2] wegen der endlichen Amplitude der Eingangsspannung und der endlichen Vormagnetisierung $\Phi_0$ zum Zeitpunkt $t = 0$ entsprechend dem Induktionsgesetz:

$$\Phi_\mathrm{W}(t) = \Phi_0 + \frac{1}{N_1} \int_0^t u_1(\tau) \cdot \mathrm{d}\tau < \infty \quad \{23\}$$

Der magnetische Fluss ist somit endlich und verteilt sich auf die Querschnittsfläche $A$ des Kerns. Bei homogener Flussaufteilung auf die Fläche lässt sich die Flussdichte berechnen zu

$$B(t) = \frac{\Phi_\mathrm{W}(t)}{A} < \infty \qquad \{24\}$$

Sie ist offensichtlich ebenfalls endlich.

---

[2] Der Index „W" steht für „Windung".

### 3.3.2. Transformationsgleichung

Die Gleichung für die Stromtransformation kann in verschiedenen Idealisierungsgraden hergeleitet werden. Bei der stärksten Idealisierung betrachtet man den Idealfall $\mu_r \to \infty$ und argumentiert folgendermaßen: Aufgrund der großen Permeabilität des Kernmaterials ergibt sich im Kern eine verschwindende magnetische Feldstärke mit

$$H(t) = \frac{B(t)}{\mu_0\mu_r} \to 0. \qquad \{25\}$$

Wendet man das Durchflutungsgesetz auf den in **Abb. 3** eingezeichneten Integrationsweg K an und berücksichtigt $H = 0$, so ergibt sich

$$\oint_K \vec{H}\cdot\mathrm{d}\vec{s} = 0 = N_1\cdot i_1(t) + N_2\cdot i_2(t) \qquad \{26\}$$

und somit

$$N_1\cdot i_1(t) = -N_2\cdot i_2(t). \qquad \{27\}$$

Gleichung $\{27\}$ vernachlässigt, dass der Primärstrom nicht nur zur Bereitstellung der sekundärseitigen Leistung, sondern teilweise auch zum Aufbau des magnetischen Flusses (Magnetisierungsstrom) und zur Speisung der spannungsgetriebenen Eisenverluste im Kern benötigt wird. Bei den Eisenverlusten handelt es sich zu wesentlichen Teilen um Wirbelstrom- und Ummagnetisierungsverluste.

Gleichung $\{27\}$ ist weder bei einem hochohmig abgeschlossenen Transformator mit $i_2 = 0$, noch bei Kernsättigung ($\vec{H} \neq \vec{0}$) anwendbar. Auf die nähere Beschreibung dieser speziellen Eigenschaften soll im Rahmen dieser Veröffentlichung jedoch verzichtet werden.

### 3.4. Leistungsübertragung

Zu Verständnis der Vorgänge beim Transformator ist es empfehlenswert, die Energieübertragung mithilfe des Poyntingvektors zu betrachten. Dieser ist definiert als das Kreuzprodukt aus elektrischer und magnetischer Feldstärke

$$\vec{S} = \vec{E} \times \vec{H} \qquad \{28\}$$

und gibt die Leistungsdichte (d. h. die Leistung pro Fläche) des elektromagnetischen Feldes an.

Da im magnetischen Kern bei idealtypischer Betrachtung $H = 0$ gilt (vgl. Gleichung $\{25\}$), verschwinden im Kern sowohl die magnetische Energiedichte $w = 1/2\, BH$, als auch die durch den Poyntingvektor beschriebene Energieflußdichte $\vec{S}$. Die Energie kann somit offensichtlich nicht durch den Kern fließen, sondern wird durch das den Kern umgebende Medium (typischerweise Luft oder Isolatoren) transportiert.

Um den Feldverlauf plausibel zu machen, soll mithilfe von **Abb. 6** der Feldverlauf in der Umgebung des Transformatorkerns qualitativ untersucht werden. Hierzu wird das Durchflutungsgesetz auf den eingezeichneten Weg $s_1$ angewendet. Da die magnetischen $H$-Feldlinien materialbedingt nahezu senkrecht aus dem Kern austreten, kann man das eingezeichnete $H$-Feld zwischen den Schenkeln unter Berücksichtigung von $H = 0$ im Kern näherungsweise mit

$$\oint_{s_1} \vec{H}\cdot\mathrm{d}\vec{s} = H\cdot l = N_1\cdot i_1 \qquad \{29\}$$

berechnen. Je größer der Primärstrom ist, umso größer ist also das $H$-Feld zwischen den Schenkeln des Transformatorkerns.

Das Vorhandensein des elektrischen Feld in der Umgebung des Kerns lässt sich mithilfe des Induktionsgesetzes erklären, indem man dieses auf geschlossenen Wegen um den Kern anwendet.

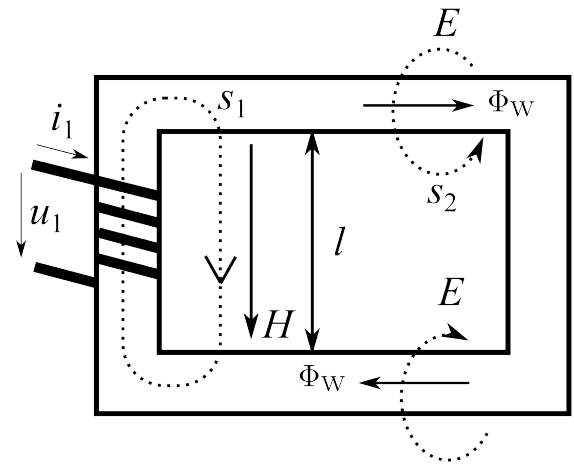

**Abb. 6:** Qualitativer Feldverlauf außerhalb des Transformatorkerns

Für den Umlaufweg $s_2$ gilt dann beispielsweise:

$$\oint_{s_2} \vec{E}\cdot\mathrm{d}\vec{s} = -\frac{\mathrm{d}\Phi_W}{\mathrm{d}t} = -\frac{u_1(t)}{N_1} \qquad \{30\}$$

Die Gleichungen zeigen, dass das **$H$-Feld vom Primärstrom $i_1$** abhängt, während das **$E$-Feld von der Primärspannung $u_1$** bestimmt wird.

Der beschriebene Zusammenhang ermöglicht es, ohne großen Aufwand den Bezug zwischen dem Poyntingvektor und der elektrischen Leistung herzustellen, denn mithilfe der Zusammenhänge ergibt sich ohne weiteren Zwischenschritt, dass die primärseitig eingespeiste Leistung $P_1 = u_1\cdot i_1$ proportional zum Betrag $S$ des Poyntingvektors $\vec{S} = \vec{E} \times \vec{H}$ ist.

Insgesamt ergibt sich bei dem dargestellten Transformator qualitativ der in **Abb. 7** dargestellte Feldlinienverlauf, wobei die Pfeilrichtungen den Fall für Einspeisung von elektrischer Energie in den Primärkreis mit $u_1 > 0$, $i_1 > 0$ darstellen. Entsprechend dem Verlauf des Poyntingvektors wird die Energie außerhalb des Kerns von der Primär- zur Sekundärseite transportiert.

Im Ergebnis bedeuten die Überlegungen, dass die i. A. als Nebensächlichkeit betrachteten Streufelder:

- magnetisches Streufeld $H$

- elektrisches Wirbelfeld $E$ um den Kern

die eigentliche Grundlage für den gesamten Energie-transport darstellen.

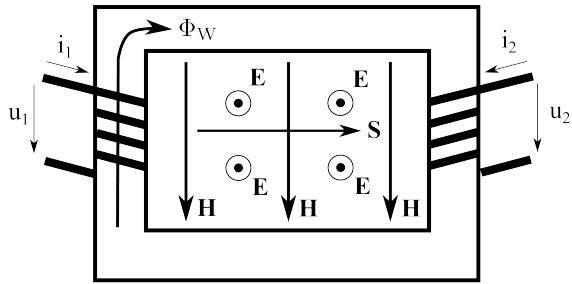

**Abb. 7:** Die Energie beim Transformator wird am Kern entlang (aber außerhalb des Kerns) von der Primär- zur Sekundärseite transportiert.

Das Verhältnis zwischen den Feldern im Kern und außerhalb des Kerns kann folgendermaßen zusammengefasst werden:

- Der Kern und die mit seiner Geometrie und seinen Materialeigenschaften verbundenen Randbedingungen lenken die Energie in die gewünschte Richtung.
- Das umgebende Medium hingegen transportiert die Energie in Form einer geführten elektromagnetischen Welle.

Weitere Erläuterungen zum Thema der Energieübertragung am Transformator, insbesondere für andere Transformatorgeometrien, finden sich in den Artikeln [9] und [10].

### 3.5. Zusammenhang zwischen Netzwerkgrößen und der Feldgrößen

Zum Verständnis des Verhältnisses zwischen den elektrischen und magnetischen Größen sollen exemplarisch die Größen an einer elektrischen Spule unter idealtypischen Bedingungen dargestellt werden:

Um mittels der Maxwellgleichungen von der Spulenspannung $u$ auf den zugehörigen Strom $i$ zu schließen, muss man in dieser Reihenfolge das Induktionsgesetz, die Materialgleichung für den magnetischen Kern und das Durchflutungsgesetz anwenden und dabei die Geometrie der Anordnung berücksichtigen (**Abb. 8**). Zwischen dem magnetischen Fluss und dem Strom einer Spule wird dabei die Beziehung $\Phi = Li$ hergestellt.

Die Induktivität $L$ der Anordnung als Spulenkonstante berechnet sich auf der Basis der Geometrie und der Materialkonstanten $\mu_r$. Dies in Gl. {17} eingesetzt, ergibt folgenden dynamischen Zusammenhang zwischen der Spannung und dem Strom einer Spule

$$u(t) = L \cdot \frac{\mathrm{d}i(t)}{\mathrm{d}t}, \qquad \{31\}$$

welcher gelegentlich unzutreffend als „Induktionsgesetz" bezeichnet wird, obwohl das Induk-tionsgesetz, das Durchflutungsgesetz und die Materialgleichungen in diese Gleichung eingehen.

$$
\begin{array}{ll}
u & \dfrac{\displaystyle\oint_{\partial A} \vec{E} \cdot \mathrm{d}\vec{s} = -\int_A \frac{\partial \vec{B}}{\partial t} \cdot \mathrm{d}\vec{A}}{\text{Induktionsgesetz}} \quad B = \Phi/A \\[2em]
i & \dfrac{\text{Durchflutungsgesetz}}{\displaystyle\oint_{\partial A} \vec{H} \cdot \mathrm{d}\vec{s} = \sum I + \int_A \frac{\partial \vec{D}}{\partial t} \cdot \mathrm{d}\vec{A}} \quad H
\end{array}
$$

Material-gleichung $B = \mu H$

**Abb. 8:** Darstellung der einander zugeordneten elektrischen und magnetischen Größen bei Spule und Transformator.

Zudem ist zu beachten, dass die Materialgleichungen und die Grundgleichungen (Induktionsgesetz und Durchflutungsgesetz) verschiedene Aussagekraft haben:

- Die Materialgleichungen beschreiben eine physikalische Vereinfachung bzw. Modellierung statischer und auch zeitlicher Prozesse im Material und sind damit klar von den universell gültigen Maxwell'schen Gleichungen zu unterscheiden.
- Das Induktionsgesetz und das Durchflutungsgesetz sind hingegen universell und zu jedem Zeitpunkt geltende Gesetzmäßigkeiten.

Somit beschreiben die elektrische Umlaufspannung an einer Spule und das Auftreten der magnetischen Flussänderung in der Spule nicht, wie manche Beschreibung suggeriert, zwei zeitlich aufeinanderfolgende physikalische Einzelvorkommnisse, sondern können gewissermaßen „synonym" verwendet werden.

### 4. Diskussion häufiger Fehldarstellungen

### 4.1. Vorzeichen der elektrischen Größen am Transformator

Ein Vergleich zahlreicher Bücher der einführenden Lehrbücher zeigt, dass sich viele Autoren hinsichtlich der Vorzeichen bei Spulen und Transformatoren irren und einige vorsorglich sogar ganz auf eindeutige Angaben zu den Vorzeichen verzichten, indem sie beispielsweise Betragsgleichungen oder Doppelpfeile verwenden. Daher sollen die bisherigen Zusammenhänge zusammengefasst werden:

*Bei einem Transformator weisen die Spannungen an Primär- und Sekundärseite gleiche Vorzeichen, die Ströme jedoch entgegengesetzte Vorzeichen auf – jeweils bezogen auf eine dem magnetischen Fluss rechtshändig zugeordnete Umlaufrichtung.*

Die Gleichheit der Spannungsvorzeichen beruht darauf, dass auf Primär- und Sekundärseite derselbe Fluss im gleichen Wicklungssinn umschlossen wird. Da für Primär- und Sekundärseite auch dasselbe Induktionsgesetz gilt, können sich die Vorzeichen nicht unterscheiden. Die Unterschiedlichkeit der Stromvorzeichen beruht darauf, dass sie wegen der Materialbedingung $H = 0$ im Kern, die mit den Strömen vorhandenen magnetischen Felder gegenseitig kompensieren müssen, so dass ein Strom im Uhrzeigersinn und der andere Strom im Gegenuhrzeigersinn um den Kern kreist.

### 4.2. Behauptung: Das Induktionsgesetz reicht zur Erklärung aus.

Leider wird nur in wenigen einführenden Lehrbüchern bei den Erklärungen zum Transformator das Durchflutungsgesetz verwendet, wie es beispielsweise in [11] (Kap. 7.6.4 – Transformator mit Eisenkern) erfolgt.

Stattdessen wird das Durchflutungsgesetz oft durch eine Kombination aus dem Energieerhaltungssatz und der Annahme, dass im Transformator keine Energie zwischengespeichert wird, ersetzt oder versteckt über Gleichungen für spezielle Anordnungen eingeführt. Derartige Ansätze lassen leider die Gelegenheit ungenutzt, an einem für Schüler und Studenten überschaubaren Beispiel das Ineinandergreifen von Induktions- und Durchflutungsgesetz zu zeigen.

Die übliche Lehrpraxis bewirkt nun zweierlei:

- Einerseits erkennen die Lernenden nur mit großer Mühe, auf welche Art und Weise die Materialgleichungen in die Rechnungen eingehen. Diese Kenntnis ist jedoch für das Verständnis insbesondere der nichtlinearen Eigenschaften sehr wichtig.

- Andererseits "verbergen" die Ansätze, dass das Durchflutungsgesetz für den Transformator eine gleich große Relevanz hat wie das Induktionsgesetz. Besonders irreführend ist in diesem Zusammenhang die gelegentlich anzutreffende Formulierung, der zufolge Ströme "induziert" würden. Dies führt bei vielen Lernenden zu der einseitigen Ansicht, der Transformator beruhe auf der elektromagnetischen Induktion allein.

Auch in vielerprobten Schulbüchern sind die Darstellungen leider noch nicht ausgereift. So heißt es beispielsweise in [12]:

„*Den Zusammenhang zwischen den Stromstärken $I_2$ und $I_1$ können wir mit dem Induktionsgesetz herleiten. Nach der Grundgleichung des Trafos ist die zeitliche Änderung des Flusses* $\dot{\Phi}(t) = U_1(t)/n_1$ *eindeutig durch $U_1$ bestimmt. Bei Belastung kann der Fluss $\Phi(t)$ deshalb durch eine Sekundärstromstärke $I_2$ nicht verändert werden. Deren Zusatzfluss* $\Phi_2(t) = \mu_0\mu_r A_2 n_2 I_2(t)/l_2$ *muss deshalb durch einen Fluss $\Phi_1(t)$ zu null ausgeglichen werden:*

$\Phi_1(t) + \Phi_2(t) = 0.$

*Damit folgt*

$\mu_0\mu_r A_1 n_1 I_1(t)/l_1 \; + \; \mu_0\mu_r A_2 n_2 I_2(t)/l_2 = 0.$"

Ein aufmerksamer Schüler wird bei dieser Darstellung kaum verstehen, dass mit $\Phi_1$ der zum primären *Zusatzstrom* gehörige Fluss gemeint ist und nicht etwa der zum primären *Gesamtstrom* gehörige Fluss. Er wird sich auch fragen, weshalb die Autoren zwar auf das Induktionsgesetz verweisen, den zur Erklärung grundlegenden Durchflutungssatz aber nicht erwähnen, obwohl nur dieser die Stromstärke explizit enthält. Zudem geht in der Darstellung die Wichtigkeit der Permeabilitätszahl $\mu_r$ verloren.

### 4.3. Behauptung: Klemmenspannung und induzierte Spannung kompensieren einander.

Die Darstellung der Größen am Transformator entsprechend **Abb. 8** erweist sich als sehr nützlich, um die teilweise recht eigenwilligen Beschreibungen in der Literatur kritisch zu beleuchten. Als ein Beispiel soll das folgende Zitat aus [14] – Kapitel 5.6.1 betrachtet werden:

"*Wird an die Primärspule $L_1$ des unbelasteten Transformators die Eingangsspannung*

$$U_1 = U_0 \cos\omega t$$

*angelegt, so wird in $L_1$ der Strom $I_1$ fließen, der einen magnetischen Fluss $\Phi_m$ erzeugt. Dieser bewirkt eine Induktionsspannung*

$$U_{ind} = -L_1\frac{dI_1}{dt} = -N_1\frac{d\Phi_m}{dt},$$

*welche der von außen angelegten Spannung $U_1$ entgegengesetzt gleich ist [...]*"

Die Beschreibung wirkt zunächst einleuchtend, erweist sich aber bei genauerer Betrachtung eher als ein umständlicher Zirkelschluss denn als eine nützliche Erklärung der physikalischen Zusammenhänge (vgl. **Abb. 9**). Ein wesentliches Problem der Darstellung besteht darin, dass die induzierte Spannung $U_{ind}$ in Wirklichkeit nicht erst nach dem Ablaufen eines komplizierten Prozesses aus der Eingangsspannung $U_1$ hervorgeht, sondern schon zum Beginn des Experimentes mit ihr identisch ist. Erst im Laufe der Zeit ergeben sich verursacht durch ohmsche Spannungsabfälle am Draht leichte Abweichungen zwischen beiden Größen.

An Gleichung {16} ist zu erkennen, dass die Klemmenspannung und die induzierte Spannung an Spulen und Transformatoren in Wirklichkeit in einem Verhältnis gänzlich anderer Art stehen als üblicherweise behauptet:

> Die induzierte Spannung steht der Klemmenspannung in Wirklichkeit nicht etwa entgegen, sondern sie enthält die Klemmenspannung als einen Teil ihrer selbst.

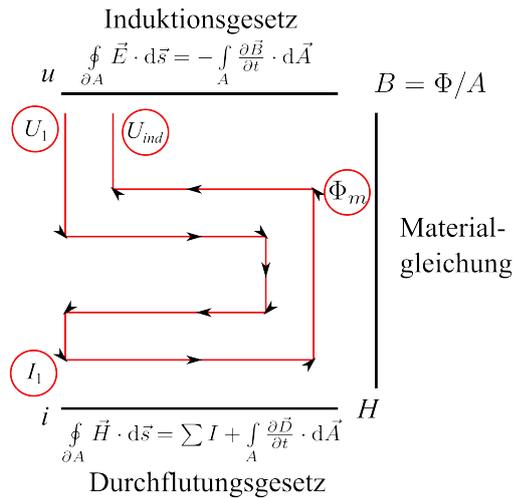

Induktionsgesetz

$$u \quad \oint_{\partial A} \vec{E} \cdot d\vec{s} = -\int_A \frac{\partial \vec{B}}{\partial t} \cdot d\vec{A} \quad B = \Phi/A$$

$U_1$  $U_{ind}$

$\Phi_m$

Material-
gleichung

$I_1$

$$i \quad \oint_{\partial A} \vec{H} \cdot d\vec{s} = \sum I + \int_A \frac{\partial \vec{D}}{\partial t} \cdot d\vec{A} \quad H$$

Durchflutungsgesetz

**Abb. 9:** Argumentationsschema zu dem Zusammenhang zwischen Klemmenspannung und induzierter Spannung nach [5]. Bei der Argumentation werden das Induktionsgesetz, die Materialgleichung für den magnetischen Kern und das Durchflutungsgesetz zweimal angewendet. Durch die jeweils doppelte Anwendung der Gleichungen werden diese nicht wirksam.

Letztlich beruht die Vorstellung, dass die Klemmenspannung durch die Induktionsspannung „kompensiert" wird, wahrscheinlich auf einem misslungenen Versuch, die elektromagnetischen Feldtheorie und die Netzwerktheorie miteinander zu kombinieren. Der Versuch, die physikalischen Vorgänge bei der Induktion mithilfe der Netzwerktheorie schlüssig zu erklären, scheiterte unserer Auffassung nach deshalb, weil die Netzwerktheorie die Gültigkeit der Kirchhoff'schen Maschengleichung postuliert, der zufolge die Summe aller Spannungen in einem geschlossenen Umlauf den Wert Null ergibt:

$$\oint_{\partial A} \vec{E} \cdot d\vec{s} = 0. \qquad \{32\}$$

Elektromagnetische Induktion liegt dem Induktionsgesetz entsprechend jedoch nur dann vor, wenn der rechte Term gerade einen von null verschiedenen Wert annimmt:

$$\oint_{\partial A} \vec{E} \cdot d\vec{s} = -\int_A \frac{\partial \vec{B}}{\partial t} \cdot d\vec{A}. \qquad \{33\}$$

Die Netzwerktheorie ist zwar in der Lage, den Zusammenhang von Leiterstrom und Klemmenspannung an induktiven Bauelementen in Form einer Differentialgleichung zu beschreiben. Zur Beschreibung modellfremder Größen wie elektrischer oder magnetischer Felder eignet sie sich jedoch nur in einem sehr unbefriedigenden Maße.

Die im Zitat implizit enthaltene Aussage, dass die an einer Spule angelegte Klemmenspannung durch die induzierte Spannung kompensiert wird, findet sich dennoch leider in vielen etablierten Lehrbüchern für das Grundstudium (z. B. [15] – Kapitel 8.3.8, [16] – Kapitel 4.3.1.4) und in zahlreichen Praktikumsanlei-

tungen an Universitäten ([17], [18], [19], [20], [21], [22]) wieder. So heißt es beispielsweise in [15]:

*„An die Klemmen der Primärspule legen wir die Spannung $U_1$. Da der Widerstand dieser Spule rein induktiv ist, wird $U_1$ kompensiert durch eine entgegengesetzt gleiche Selbstinduktionsspannung, die ihrerseits nach (8.3) gleich der Flussänderung durch alle $N_1$ Windungen ist [...]"*

Durch derartige Erklärungen werden die Lernenden letztlich vor schwerwiegende Widersprüche gestellt:

- Der Aufbau des magnetischen Flusses beruht angeblich auf dem Gleichgewicht zwischen Induktionsspannung und Klemmenspannung. Doch weshalb sollte eine Veränderung des Flusses gerade auf einem Gleichgewicht, d. h. der Voraussetzung für Nichtveränderung, beruhen?

- In einer elektrischen Spule entsteht angeblich eine induzierte Spannung. Die Materialbedingung für eine ideale Spule lautet jedoch $E = 0$ und somit $U = 0$.

### 4.4. Behauptung: Beim Transformator fließt die übertragene Energie durch den Transformatorkern.

Gelegentlich findet sich auch die Vorstellung, dass die Kerne von Transformatoren in ihrer Funktion als Energiezwischenspeicher genutzt würden. So heißt es beispielsweise in [23], Kap. 12.1:

*„Ein Transformator [...] ist ein elektromagnetischer Energiewandler, der in seiner Primärwicklung [...] elektrische Energie in magnetische Energie umwandelt; in der Sekundärwicklung [...] wird die magnetische Energie wieder in elektrische Energie umgewandelt."*

Ähnliche Ausführungen finden sich auch in [11] – Kap. 7.6.4 und [25] – Kap. 4.9.1.1.

Wie in Kapitel 3.4 erläutert wurde, ist beim idealtypischen Transformator der Kern mit $H = 0$ vollkommen frei von elektromagnetischer Feldenergie, da die Energiedichte verschwindet. Ebenso ist die durch den Poyntingvektor beschriebene Leistungsdichte im Transformatorkern gleich null. Bei Anordnungen mit realen Transformatoren werden diese Bedingungen immer noch in guter Näherung eingehalten. Transformatoren werden daher i. A. nicht als Energiespeicher genutzt, und auch die Energieübertragung findet überall anders als im Kern statt.

Eine nennenswerte Zwischenspeicherung von Energie kommt allenfalls bei Transformatorkernen mit Luftspalt oder bei elektrischen Motoren in Betracht. Die Energie konzentriert sich dabei auf das Volumen des Luftspaltes.

#### 4.5. Behauptung: Magnetische Streufelder sind nebensächlich

Es fällt auf, dass den Streufeldern in der Literatur oft undifferenziert die Bedeutung einer „Störgröße" zukommt. So heißt es beispielsweise in [11] – Kapitel 7.1.6 (Belasteter Transformator):

*„Der in der Sekundärspule induzierte Strom $i_2$ wirkt nach der Lenz'schen Regel seiner Ursache, also dem vom Strom $i_1$ erzeugten Fluss $\Phi_1$, entgegen, indem der von $i_2$ erzeugte Fluss $\Phi_2$ dem Fluss $\Phi_1$ entgegengerichtet ist. Beim Experimentiertrafo wird dadurch der Fluss $\Phi_1$ zum Teil aus der Spule gedrängt; es entsteht Streufluss. Damit ist die feste magnetische Kopplung zwischen Primär- und Sekundärspule aufgehoben und die Funktion des Transformators beeinträchtigt".*

Wie in Abschnitt 3.4 erläutert wird, ist es nicht richtig zu behaupten, dass die Funktion des Transformators durch die Streufelder „beeinträchtigt" würde. Denn die Streufelder sind in Wirklichkeit die Träger von Energie und Information; erst durch sie wird die Funktion des Transformators überhaupt erst ermöglicht. Richtig ist hingegen, dass sich ein realer Transformator mit großem Streufluss nur noch näherungsweise mit den Gleichungen für den idealen Transformator beschreiben lässt.

#### 4.6. Behauptung: Je größer der Laststrom, umso eher sättigt der Transformator

Weit verbreitet ist auch die Annahme, dass Lastströme zur Sättigung des Transformators beitragen. Diesem Eindruck möchten wir im Folgenden widersprechen.

Um den Vorgang der Sättigung zu verstehen, denken wir uns ein hysteresebehaftetes Material entsprechend **Abb. 10** in einem leerlaufenden Transformator (bzw. einer elektrischen Spule) eingebaut. Als Vorbemerkung sei zudem nochmals erwähnt, dass die Flussdichte $B$ mittels des Induktionsgesetzes grundsätzlich mit einer elektrischen Spannung in einem materialunabhängigen Verhältnis steht, während die magnetische Feldstärke $H$ über das Durchflutungsgesetz mit elektrischen Strömen verknüpft ist (vgl. **Abb. 8**).

Da vor dem Einschalten des Transformators alle elektrischen Ströme gleich Null sind, gilt im Anschaltmoment im Kernmaterial die Beziehung $H = 0$. Der Magnetisierungsvorgang beginnt demzufolge (mit Ausnahme von fabrikneuem Material) auf einem der beiden Remanenzpunkte bei $H = 0$. Wir gehen im Folgenden davon aus, dass der Magnetisierungsvorgang auf dem oberen Remanzpunkt beginnt.

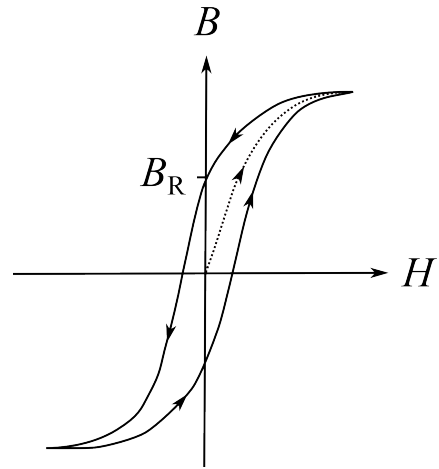

**Abb. 10:** Hysteresekurve eines weichmagnetischen Transformatorkerns mit der Remanenzflussdichte $B_R$.

Speisen wir nun primärseitig in den (leerlaufenden) Transformator eine vorgegebene Wechselspannung $u_1(t)$ ein, so wissen wir aus dem Induktionsgesetz, dass damit materialunabhängig eine Veränderung des magnetischen Flusses und somit des $B$-Feldes im Kern einhergeht, denn es gilt:

$$\Phi_W(t) = \Phi_0 + \frac{1}{N_1} \int_0^t u_1(\tau)\mathrm{d}\tau \qquad \{34\}$$

Angenommen, $u_1(t)$ ist ein sinusförmiges Signal der Frequenz $f = 1/T$ mit nur einer einzigen Periode. Dann wird sich der magnetische Fluss während der ersten Sinushalbwelle um den Betrag

$$\Delta\Phi = \frac{1}{N_1} \int_0^{T/2} U_1 \cdot \sin(2\pi f t)\mathrm{d}t = \frac{U_1}{N_1 f \pi} \quad \{35\}$$

vergrößern und während der zweiten Sinushalbwelle um denselben Betrag wieder verringern[3].

Das Problem eines steigenden $B$-Feldes besteht darin, dass entsprechend **Abb. 8** das $H$-Feld (und damit materialunabhängig verbunden der elektrische Primärstrom – der Sekundärstrom ist ja voraussetzungsgemäß gleich Null) ab einer gewissen Stärke der Vormagnetisierung stark überproportional anwächst. Das starke Anwachsen des $H$-Feldes und des Stromes bei nur geringen Flussänderungen nennt man Sättigung.

---

[3] Aus Gleichung {34} folgt, dass bei $\Phi_0 \neq 0$ und gleichanteilsfreiem $u_1(t)$ beim Betrieb eines Transformators im zeitlichen Mittel eine mittlere Flussdichte von $\Phi_0$ herrschen sollte. Es scheint also so, als würde der Transformatorkern normalerweise unsymmetrisch ausgesteuert werden. Das stimmt jedoch in Wirklichkeit nicht. Eine wesentliche Ursache hierfür ist der ohmsche Widerstand der Spulenwicklungen.
Zum Verständnis müssen wir uns vergegenwärtigen, dass entsprechend Gleichung {16} ein ohmscher Spannungsabfall am Spulendraht die Induktionsspannung, die für den Aufbau des magnetischen Flusses aufgewendet wird, verringert. Da bei einem positiven Offset $\Phi_0 > 0$ der magnetischen Flussdichte im Mittel ebenfalls ein positiver Offset des Stromes (verursacht durch die abknickende magnetische Kennlinie) vorliegt, ist der durch die positive Sinushalbwelle verursachte aufmagnetisierende Flusshub $\Delta\Phi^+$ etwas kleiner als der durch die als negative Sinuswelle verursachte abmagnetisierende Flusshub $\Delta\Phi^-$. Ein möglicher Offset der magnetischen Flussdichte baut sich auf diese Weise typischerweise innerhalb nur weniger Sinushalbwellen von alleine ab.

Wie in Gleichung {35} zu erkennen ist, sind die entscheidenden Parameter, die zur Sättigung führen, die Amplitude $U_1$ der Eingangsspannung, die Frequenz $f$, die Windungszahl $N_1$ und der Querschnitt $A = \Phi/B$ des Kerns. Ströme kommen in der Gleichung jedoch nicht vor.

Der wesentliche Grund, weshalb der sekundärseitige Laststrom keinen Beitrag zur Sättigung liefert, besteht darin, dass das von ihm erzeugte magnetische Feld vom sogenannten primären Zusatzstrom kompensiert wird. Tendenziell führen hohe Lastströme sogar zu einer Verringerung der Sättigung, da die Lastströme ohmsche Spannungsabfälle (in der idealtypischen Betrachtung vernachlässigt) an den Wicklungen von Primär- und Sekundärseite erzeugen und dadurch ein geringerer Teil der Primärspannung zum Flussaufbau verwendet wird. Ebenfalls vorstellbar ist, dass bei großen Lastströmen ein Teil der Spannung am Innenwiderstand der Quelle abfällt und gleichermaßen eine Verringerung der induzierten Spannung sowie der Sekundärspannung hervorruft.

### 4.7. Behauptung: Eine hohe Wicklungszahl an der Primärseite eines Transformators führt zu einem großen magnetischen Fluss im Kern

Im Physikunterricht in der Schule wird häufig die magnetische Flussdichte in einer langen Spule untersucht. So heißt es beispielsweise in [24] – Kap. 6.2.1:

*„Die magnetische Feldstärke[4] B des homogenen Feldes im Innern einer langen Spule ist proportional zur Stromstärke I und zur Windungszahl n/l. Mit der magnetischen Feldkonstante $\mu_0$ berechnet sich B zu:*

$$B = \mu_0 I \, \frac{n}{l}\text{"}$$

Dieses Ergebnis übertragen viele Studenten unbedacht auf *spannungsgespeiste* Spulen und Transformatoren und denken, dass die Wicklungszahl auch hier den magnetischen Fluss erhöht. Wie Gleichung {35} zeigt, ist jedoch das Gegenteil der Fall:

- *Je größer die Wicklungszahl $N_1$ ist, umso kleiner ist der magnetische Wechselfluss im Kern und umso geringer ist die Sättigung.*

Eine Erhöhung des magnetischen Flusses durch höhere Wicklungszahlen findet somit ausschließlich bei *stromgespeisten* Spulen und Transformatoren statt, die jedoch praktisch kaum verwendet werden.

---

[4] Im Metzler heißen in Anlehnung an internationale Gepflogenheiten $B$ – magnetische Feldstärke und $H$ – magnetische Erregung. Wir verwenden die in Deutschland üblichen Bezeichnungen $B$ – magnetische Flussdichte und $H$ – magnetische Feldstärke.

### 4.8. Behauptung: Bei Transformatoren mit Luftspalt verringert sich der Kopplungsfaktor, weil der Luftspalt die Streuinduktivität vergrößert.

Es ist bekannt, dass mithilfe eines Luftspaltes im Transformatorkern die Kopplung beider Transformatorseiten verringert werden kann.

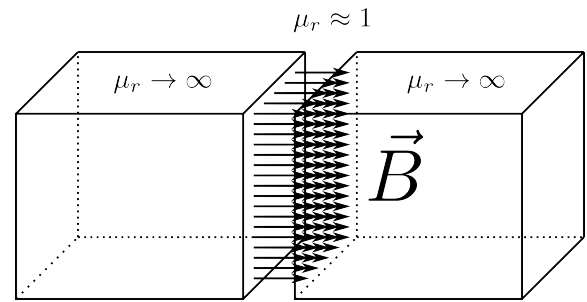

**Abb. 11:** Ein Luftspalt in einem Transformatorkern erzeugt selbst keinen zusätzlichen Streufluss, da die magnetischen Flusslinien wie dargestellt senkrecht aus dem Kern austreten und auf dem gegenüberliegenden Pol senkrecht wieder eintreten. Der erhöhte Streufluss ergibt sich indirekt. Der Grund ist die geringere Induktivität der Anordnung und die dadurch verursachte Vergrößerung des Magnetisierungsstroms.

Viele Studierende denken jedoch, dass die Feldlinien im Bereich des Luftspaltes „abbiegen" und den durch den Kern vorgegebenen Magnetkreis verlassen. Das ist jedoch in Wirklichkeit nicht zutreffend: Denn entsprechend **Abb. 11** tritt aufgrund der hohen Permeabilität des magnetischen Materials der magnetische Fluss i. A. mit einem Winkel von näherungsweise 90° aus dem Material aus und tritt bei den typischerweise nur wenige Millimeter dicken Luftspalten unter dem gleichen Winkel wieder in den gegenüberliegenden Pol ein.

Wir wollen vereinfachend einen unbelasteten Transformator (d. h. eine Spule) untersuchen. Die wesentliche Veränderung, die durch das Anbringen eines Luftspaltes im Transformatorkern erfolgt, ist nicht die Erhöhung der Streuinduktivität, sondern die Verringerung der Gesamtinduktivität der Anordnung. Damit verbunden ist, dass zum Aufbau des gleichen magnetischen Gesamtflusses ein höherer Strom (der sogenannte Magnetisierungsstrom) fließen muss als bei einem Kern ohne Luftspalt. Mithilfe von **Abb. 6** und Gleichung {29} wird klar, dass es letztlich die mit dem Primärstrom verknüpfte Vergrößerung des $H$-Feldes zwischen den Transformatorschenkeln ist, die den Streufluss vergrößert.

Der Streufluss vergrößert sich demzufolge entlang des gesamten Transformatorkerns, und nicht nur am Ort der Kernunterbrechung!

### 5. Zusammenfassung

Es wurde gezeigt, dass die Darstellung der physikalischen Vorgänge bei Spule und Transformator in vielen Bücher als kritisch anzusehen ist, da oft die gegebenen Erklärungen in deutlichem Widerspruch zu den physikalischen Gesetzen stehen.

Mit Hilfe des Induktionsgesetzes, des Ampere'schen Durchflutungsgesetzes und der Materialgleichungen lässt sich ein konsistentes Bild der Vorgänge zeichnen. Das direkt aus dem Induktionsgesetz gewonnene Spannungsverhältnis am Transformator ist bei gleichem Wicklungssinn der Spulen, entgegen der häufigen Darstellung in der Literatur, positiv. D. h. Primär- und Sekundärspannung sind phasengleich. Das Stromverhältnis kann direkt aus dem Ampere'schen Durchflutungsgesetz gewonnen werden, wobei der Primär- und Sekundärstrom 180° phasenverschoben sind. Es wurde versucht, gängige und oft falsche Darstellungen zum Verhalten eines Transformators richtigzustellen. Wichtige Resultate waren dabei der Energiefluss entlang, aber außerhalb des Kerns, die Bedeutung der Streufelder bei der Energieübertragung und ein anschauliches Verständnis des Prozesses der Sättigung eines Transformators. Wir möchten die Lehrbuchautoren, Vorlesenden, Lehrer und Praktikumsleiter dazu einladen, die gängigen Darstellungen zu Spule und Transformator gründlich zu überdenken, um den Schülern und Studenten den einen oder anderen unnützen Stolperstein aus dem Weg zu räumen, der bislang liegengeblieben ist.

## 6. Danksagung

### Literatur

[1] Maxwell, James Clerk (1865): A Dynamical Theory of the Electromagnetic Field. In: Royal Society Transactions 155, S. 459–512

[2] Kapp, Gisbert: Transformatoren für Wechselstrom und Drehstrom – Eine Darstellung ihrer Theorie, Konstruktion und Anwendung, Julius Springer, Berlin, 1907.

[3] Fließbach, Torsten: Elektrodynamik – Lehrbuch zur Theoretischen Physik II, 5. Auflage, Spektrum akademischer Verlag, Heidelberg, 2008.

[4] Kröger, Roland; Unbehauen, Rolf: Zur Theorie der Bewegungsinduktion. In: Archiv für Elektronik und Übertragungstechnik 1982, S. 359–366

[5] Fakultät Physik, TU Wien, Skript zur Theoretischen Physik gelesen von Prof. Gerhard Ecker

[6] Flanders, H (1973): Differentiation under the Integral Sign, In: American Mathematical Monthly 80, S. 615–627

[7] Haus, Herman A; Electromagnetic fields and Energy, Kap. 8.4, Internetlink

[8] Simonyi, K.: Theoretische Elektrotechnik, 5. Auflage, VEB Deutscher Verlag der Wissenschaften, Berlin, 1973

[9] Herrmann, F.; Schmid, G. Bruno (1986): The Poynting vector field and the energy flow within a transformer. In: American Journal of Physics 54, 6, S. 528–531

[10] Edwards, J., Saha, T. K.: Power flow in transformers via the poynting vector. Homepage der Queensland University of Technology: Edwards00.pdf (Stand Feb. 2011)

[11] Bausch, H., Steffen, H.: Elektrotechnik – Grundlagen, 5. Auflage, Wiesbaden, Teubner, 2004.

[12] Schwab, Adolf: Begriffswelt der Feldtheorie, 4. Auflage, Kap. 3, 1993

[13] Bader, F.; Dorn F.: Physik Sek II – Gymnasium Gesamtband, Schroedel, Braunschweig, 2000.

[14] Demtröder, Wolfgang: Experimentalphysik 2 – Elektrizität und Optik, 5. Auflage, Springer, Berlin, Heidelberg, 2009.

[15] Meschede, Dieter: Gerthsen Physik, 24. überarbeitete Auflage, Springer, Berlin, 2010

[16] Frohne, Heinrich; Löcherer, Karl-Heinz; Müller, Hans; Harriehausen, Thomas; Schwarzenau, Dieter: Möller – Grundlagen der Elektrotechnik, 21. Auflage, Vieweg + Teubner, Wiesbaden, 2008.

[17] Praktikumsunterlagen Transformator Universität-GH Essen (Stand: 08.05.2011)

[18] Praktikumsunterlagen Transformator Universität Ulm (Stand: 08.05.2011)

[19] Praktikumsunterlagen Wechselstromkreis Universität Greifswald (Stand: 08.05.2011)

[20] Praktikumsunterlagen Wechselstromkreis Universität Tübingen (Stand: 08.05.2011)

[21] Praktikumsunterlagen Transformator Technische Universität Dresden (Stand: 08.05.2011)

[22] Praktikumsunterlagen Transformator Technische Universität Graz (Stand: 08.05.2011)

[23] Flegel, Georg; Birnstiel, Karl; Nerreter, Wolfgang: Elektrotechnik für Maschinenbau und Mechatronik, 8. Auflage, Hanser, München, 2004.

[24] Grehn, J.; Krause, J (Hrsg.): Metzler Physik, 4. Auflage, Schroedel, Braunschweig, 2009.

[25] Friesecke, Andreas: Die Audio-Enzyklopädie – Ein Nachschlagewerk für Tontechniker, K. G. Saur Verlag, München, 2007.